\documentclass[aps,twocolumn,prl,reprint,amsmath,amssymb,showpacs,superscriptaddress]{revtex4-2}

\usepackage{graphicx}
\usepackage{dcolumn}
\usepackage{bm}
\usepackage{amsmath,amsfonts,amsthm}
\usepackage{color}
\usepackage{mathrsfs}
\usepackage{float}
\usepackage{ulem}
\usepackage{txfonts}
\usepackage{wasysym}
\usepackage{cancel}

\usepackage{hyperref}
\hypersetup{colorlinks=true,linktoc=all,linkcolor=blue,breaklinks=true,citecolor=blue,urlcolor=blue}

\AtBeginDocument{%
    \newwrite\bibnotes
    \def\bibnotesext{Notes.bib}
    \immediate\openout\bibnotes=\jobname\bibnotesext
    \immediate\write\bibnotes{@CONTROL{REVTEX42Control}}
    \immediate\write\bibnotes{@CONTROL{%
    apsrev42Control,author="08",editor="1",pages="1",title="0",year="1"}}
     \if@filesw
     \immediate\write\@auxout{\string\citation{apsrev42Control}}%
    \fi
}%

\makeatletter

\newcommand{\Rmnum}[1]{\expandafter\@slowromancap\romannumeral #1@}
\makeatother

\newcommand{\bea}{\begin{eqnarray}}
\newcommand{\eea}{\end{eqnarray}}
\newcommand{\beq}{\begin{equation}}
\newcommand{\eeq}{\end{equation}}

\newcommand{\kBT}{k_\text{B}T}

\definecolor{dark-green}{rgb}{0,0.392,0}

\begin{document}
\title{Photonic heat transport through a Josephson junction in a resistive environment}
\author{A. Levy Yeyati}
\affiliation{Departamento de F\'isica Te\'orica de la Materia Condensada, Universidad Aut\'onoma de Madrid, 28049 Madrid, Spain\looseness=-1}
\affiliation{Condensed Matter Physics Center (IFIMAC), Universidad Aut\'onoma de Madrid, 28049 Madrid, Spain\looseness=-1}
\affiliation{Instituto Nicol\'as Cabrera, Universidad Aut\'onoma de Madrid, 28049 Madrid, Spain\looseness=-1}
\author{D. Subero}
\affiliation{PICO Group, QTF Centre of Excellence, Department of Applied Physics,
Aalto University School of Science, P.O. Box 13500, 0076 Aalto, Finland}
\author{J. P. Pekola}
\affiliation{PICO Group, QTF Centre of Excellence, Department of Applied Physics,
Aalto University School of Science, P.O. Box 13500, 0076 Aalto, Finland}
\author{R. S\'anchez}
\affiliation{Departamento de F\'isica Te\'orica de la Materia Condensada, Universidad Aut\'onoma de Madrid, 28049 Madrid, Spain\looseness=-1}
\affiliation{Condensed Matter Physics Center (IFIMAC), Universidad Aut\'onoma de Madrid, 28049 Madrid, Spain\looseness=-1}
\affiliation{Instituto Nicol\'as Cabrera, Universidad Aut\'onoma de Madrid, 28049 Madrid, Spain\looseness=-1}

\begin{abstract}
    Motivated by recent experiments [Subero et. al. Nature Comm. {\bf 14}, 7924 (2023)] we analyze photonic heat transport through a Josephson junction in a dissipative environment. For this purpose we derive general expressions for the heat current in terms of non-equilibrium Green functions for the junction coupled in series or in parallel with two environmental impedances at different temperatures. We show that even on the insulating side of the Schmid transition the heat current is sensitive to the Josephson coupling exhibiting an opposite behavior for the series and parallel connection and in qualitative agreement with experiments. We also predict that this device should exhibit heat rectification properties and provide simple expressions to account for them in terms of the system parameters.   
\end{abstract}

\maketitle

{\it Introduction.---} The physics of Josephson junctions (JJs) has attracted great interest for many decades
\cite{tafuri_book}. Even the apparently simple case of a single JJ in a resistive environment provides a rich playground to study the interplay of quantum tunneling and dissipation \cite{caldeira_quantum_1983}, which continues to be explored until today \cite{leppakangas:2013,Jebari2018}. For this system a transition between a superconducting and an insulating phase was predicted to appear as a function of the environmental resistance ($R$), regardless of the Josephson coupling energy ($E_J$). This is the so-called Schmid-Bulgadaev (SB) transition \cite{Schmid1983,Bulgadaev1984}, which has been the object of intense theoretical~\cite{masuki_absence_2022,Sepulcre2023,Altimiras2023,Kashuba2023,giacomelli2023emergent} and experimental debate in recent years \cite{Murani2020,Hakonen2021,*Murani2021,kuzmin_observation_2023}. 

In spite of this intense activity little is known regarding this system properties beyond dc charge transport. In this respect, heat transport~\cite{pekola_colloquium_2021} can provide additional insights. In a recent experiment, Subero {\it et al.} \cite{subero_bolometric_2022} explored heat transport through a JJ in the supposedly insulating regime ($R > R_Q = h/4e^2$). Even when charge transport could be well described by dynamical Coulomb blockade theory, it was found that heat transport was sensitive to the $E_J$ value (tunable through a magnetic flux in a SQUID configuration) indicating inductive response at high frequencies. The experimental results were fitted using a phenomenological theory including an inductive term to the junction effective impedance. However, such model is incompatible with the mentioned dc charge transport properties and apparently in conflict with a SB transition.

\begin{figure}[b]
    \centering
    \includegraphics[width=\linewidth]{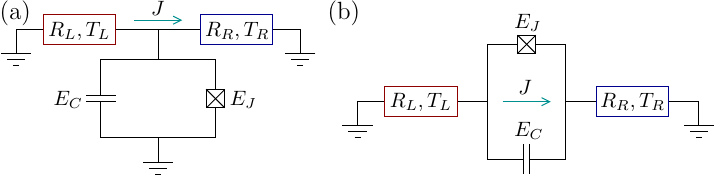}
    \caption{Schematics of the configurations considered, consisting of two resistors, $R_L$ and $R_R$, connected in (a) parallel and (b) series to a Josephson junction (defined by a Josephson, $E_J$, and a charging energy $E_c$), and holding different temperatures, $T_L$ and $T_R$, that lead to a heat current $J$.\label{fig1}}
\end{figure}

In the present work we analyze this problem from a microscopic perspective. For this purpose we develop a non-equilibrium Green functions (GFs) approach which takes into account interaction effects due to finite $E_J/E_c$ values, where $E_c$ is the junction charging energy. We find that even for $R > R_Q$, heat transport is sensitive to the Josephson coupling, in qualitative agreement with experiments. Moreover, we find that the heat current response is radically different depending on whether the junction is connected in parallel or in series to the source and drain leads. We also show that signatures of the SB transition in the heat conduction appear at sufficiently low temperatures. We finally demonstrate that asymmetric devices exhibit heat rectification properties, which provide an alternative way to detect the SB transition.  

In Fig. \ref{fig1} we schematically represent the two situations to be considered: in panel \ref{fig1}(a) the JJ is connected in parallel with the left and right resistors ($R_{L}$ and $R_R$) which are held at temperatures $T_L$ and $T_R$ respectively, whereas in panel \ref{fig1}(b) these circuit elements are connected in series, corresponding to the experimental configuration in Ref.~\cite{subero_bolometric_2022}. The aim of the theory is to determine the resulting heat current $J$ as a function of the model parameters in these two situations, assuming that temperatures are low enough to neglect excited quasiparticles in the superconducting leads. 

{\it Parallel configuration.---} Let us start with the parallel configuration~\cite{thomas_photonic_2019}. This is modeled by the following Hamiltonian
\begin{eqnarray}
    H &=& E_c N^2 - E_J \cos \varphi - \sum_{j,k} \left[ \lambda_{j,k} \left(a_{jk} + a^{\dagger}_{jk}\right) \varphi - \frac{\lambda^2_{jk}}{\hbar \omega_{jk}} \varphi^2 \right] \nonumber \\ 
   && + \sum_{j,k} \omega_{jk} a^{\dagger}_{jk} a_{jk} \;,
   \label{caldeira-leggett}
\end{eqnarray}
where $j \equiv L,R$ and the dissipative terms satisfy
\[ \frac{\pi}{2} \sum_k\lambda_{jk}^2 \delta(\omega - \omega_{jk}) =    \frac{\hbar^2 \omega}{4e^2} \mbox{Re}\left[\frac{1}{Z_j(\omega)} \right] \;, \]
$Z_j(\omega)$ being the lead impedances, which in our case correspond to pure resistances $R_j$. In this configuration, only one superconducting phase variable, $\varphi$, conjugate to the charge which is transferred through the junction, $N$, plays a role.  

Following Ref.~\cite{ojanen_mesoscopic_2008}, the heat current through lead $j$ can then be computed as
\begin{eqnarray}    
J_j &=& \langle \dot{H}_j \rangle = -i \sum_{k}  \lambda_{jk} \omega_{jk} \left[\langle a_{jk}(t) \varphi(t) \rangle - \mbox{h.c.} \right] \nonumber \\
&=& 2 \sum_{k} \lambda_{jk} \omega_{jk} \mbox{Re} G^{+-}_{a_{jk},\varphi}(t,t) \;,
\end{eqnarray}
where $G^{\alpha\beta}_{A,B}(t,t') = -i \langle T_{\cal C} B(t') A(t) \rangle$ denotes the Keldysh $\alpha\beta$ GF associated with operators $A,B$.  Using the equation of motion method and transforming it into frequency representation one can express  $G^{+-}_{a_{jk},\varphi}$ as 
\begin{equation}
G^{+-}_{a_{jk},\varphi}(\omega) = \lambda_{jk} \left[D^r(\omega) g^{+-}_{a_{jk}}(\omega) + D^{+-}(\omega) g^a_{a_{jk}}(\omega) \right]  \;,
\end{equation}
where the $g^{+-,a}_{a_{jk}}$ correspond to the uncoupled leads and, for simplicity, we denote $D^{\alpha\beta,r,a} \equiv G^{\alpha\beta,r,a}_{\varphi\varphi}$ (the superscripts $r,a$ indicate the retarded, advanced components). In this way we can express the heat current as
\begin{equation}
J_j = 2 \sum_{k} \lambda_{jk}^2 \omega_{jk} \int \frac{d\omega}{2\pi} \mbox{Re} \left[ D^r(\omega) g^{+-}_{a_{jk}} (\omega) + D^{+-}(\omega) g^a_{a_{jk}}(\omega) \right] \;.
\label{JL}
\end{equation} 

We can further use
$g^{+-}_{a_{jk}} (\omega) = -2\pi i n_j(\omega_{jk}) \delta(\omega - \omega_{jk})$ and
$g^{a}_{a_{jk}} (\omega) = 1/(\omega - \omega_{jk} - i 0^+)$, 
where $n_j(\omega)$ denotes the Bose function for the modes in the $j$-lead. Replacing these in Eq.~\eqref{JL} and taking the real part we obtain
\begin{eqnarray}
J_j {=}{\int} \frac{d\omega}{2\pi} \frac{\hbar^2\omega^2}{e^2}\mbox{Im}\left[ D^r(\omega)\coth\left(\frac{\omega\beta_j}{2}\right) {-} \frac{D^K(\omega)}{2}\right] \mbox{Re}\left[\frac{1}{Z_j(\omega)}\right] \;, \nonumber\\ 
\label{heat-keldysh}
\end{eqnarray}
where we have introduced the Keldysh GFs $D^K = D^{+-}+D^{-+}$ in the triangular representation \cite{Keldysh,*keldysh_diagram_1965} and $\beta_{j} = 1/k_BT_{j}$.
Further simplification is obtained if, as in our case, the lead admittances $1/Z_j(\omega) \equiv 1/R_j$ are frequency independent. Then, imposing heat current conservation $J_R = -J_L = J$ one gets~\cite{ojanen_mesoscopic_2008}
\begin{eqnarray}
J = -\frac{\hbar^2}{2\pi e^2} \int d\omega \frac{\omega^2 \mbox{Im} D^r(\omega)}{R_L+R_R} 
\left[n_L(\omega) - n_R(\omega)\right] \;,
\label{heat-MW}
\end{eqnarray}
which is the heat transport analog of Meir-Wingreen formula for mesoscopic charge transport \cite{meir_landauer_1992}, allowing us to define an effective heat transmission coefficient for the parallel configuration
\begin{equation}
    \tau^{\rm eff}_{\rm parallel}(\omega) = -\frac{4 \eta_L \eta_R \omega \mbox{Im} D^r(\omega)}{\eta_L + \eta_R} \;, 
    \label{effective-transmission-parallel}
\end{equation}
where $\eta_j = R_Q/(2\pi R_j) \equiv \alpha_j/2\pi$. One can further check that for the perfect-transmission case one gets $J=J_{\rm max} \equiv {\pi k_B^2(T_L^2 -T_R^2)/(12\hbar)}$ as it corresponds to a ballistic channel~\cite{pendry_quantum_1983}.

{\it Perturbation theory.---} The problem is thus mapped into the evaluation of the Keldysh GFs $D^{r,a,K}(\omega)$. In the present work we focus on the regime $R_{L,R} \gtrsim R_Q$ and $E_J < E_c$ for which perturbation theory in $E_J$ should be valid and which should be appropriate to describe the experimental results of Ref. \cite{subero_bolometric_2022}. To zeroth order in $E_J$ the GFs are given by
\begin{gather}
\begin{aligned}
D_0^{r,a}(\omega) &= \frac{1}{m \omega^2 \pm i \omega \left(\eta_L+\eta_R\right) - m \omega_c^2},\\
D_0^K(\omega) &= \frac{-2i \omega \left(\eta_L (1 + 2 n_L(\omega)) + \eta_R (1  + 2 n_R(\omega)) \right)}{\left(m (\omega^2-\omega_c^2)\right)^2 + \left(\omega (\eta_L+\eta_R)\right)^2} \;,
\label{zero-order-GF-wc}
\end{aligned}
\end{gather}
where $m = \hbar/2E_c$ and $\omega_c$ is a small frequency cutoff which warrants convergence in the Fourier transforms. At zero-th order in $E_J$ and $\omega_c\rightarrow 0$, from (\ref{effective-transmission-parallel}) we obtain  
\begin{equation}
\tau^0_{\rm parallel}(\omega) = \frac{4 \eta_L \eta_R}{(m \omega)^2 + (\eta_L+\eta_R)^2} \;,
\label{parallel-coefficient}
\end{equation}
which reaches a maximum value $4\eta_L\eta_R/(\eta_L+\eta_R)^2$ at zero frequency. 

Higher order corrections can be introduced through the self-energies $\Sigma^{r,a,K}(\omega)$ associated to the $E_J$ term in the Hamiltonian, which would allow us to evaluate the needed GFs as
\begin{eqnarray}
D^{r,a}(\omega) &=& \left[ \left(D_0^{r,a}(\omega)\right)^{-1} - \Sigma^{r,a}(\omega)\right]^{-1} \nonumber\\
D^{K}(\omega) &=& \left[1 + D^r(\omega)\Sigma^r(\omega)\right]D^K_0(\omega)\left[1+ \Sigma^a(\omega)D^a_0(\omega)\right] \nonumber\\
&& + D^r(\omega) \Sigma^{K}(\omega) D^a(\omega).
\label{dressedGFs}
\end{eqnarray}  
To lowest order in $E_J$ we have \cite{eckern_quantum_1987}
\begin{equation}
\Sigma^{r,a(1)}(\omega) = E_J \exp\left[-\frac{i}{4}D_0^{K}(t=0)\right] \rightarrow 0  \;,
\label{Hartree}
\end{equation} 
since $\mbox{Im} D_0^{K}(t=0) \rightarrow -\infty$ for $\omega_c \rightarrow 0$. One also obtains $\Sigma^{K(1)}(\omega) = 0$ as it corresponds to an effective static potential. To get the frequency-dependent corrections
it is then necessary to go to higher order. To second order in $E_J$ we find~\cite{eckern_quantum_1987}
\begin{eqnarray}
\Sigma^{r(2)}(t) &=& E_J^2 \sin \frac{D_0^r(t)}{2} \exp \left[\frac{i}{2}\left(D_0^K(t) -D^K_0(0)\right)\right] - B \delta(t) \nonumber\\
\Sigma^{K(2)}(t) &=& -iE_J^2 \cos \frac{D_0^r(t)}{2} \cos \frac{D_0^a(t)}{2} \exp \left[\frac{i}{2}\left(D_0^K(t) -D^K_0(0)\right)\right] \;, \nonumber\\
\label{sigK}
\end{eqnarray}
where $B$ is a constant such that $\Sigma^{r(2)}(\omega=0)=0$. 

From these expressions one can obtain numerical results for the self-energies \cite{SM}. In addition, fully analytical results for the Keldysh self-energies can be obtained in the limit $(R_L+R_R)/R_Q \rightarrow 1$ and $T_j \rightarrow 0$~\cite{SM} (see also comments below). Let us also mention that this perturbative analysis leads to the same predictions as dynamical Coulomb blockade theory that were used in Ref. \cite{subero_bolometric_2022} for fitting the IV curves \cite{SM}.

\begin{figure}[t]
\includegraphics[width=\columnwidth]{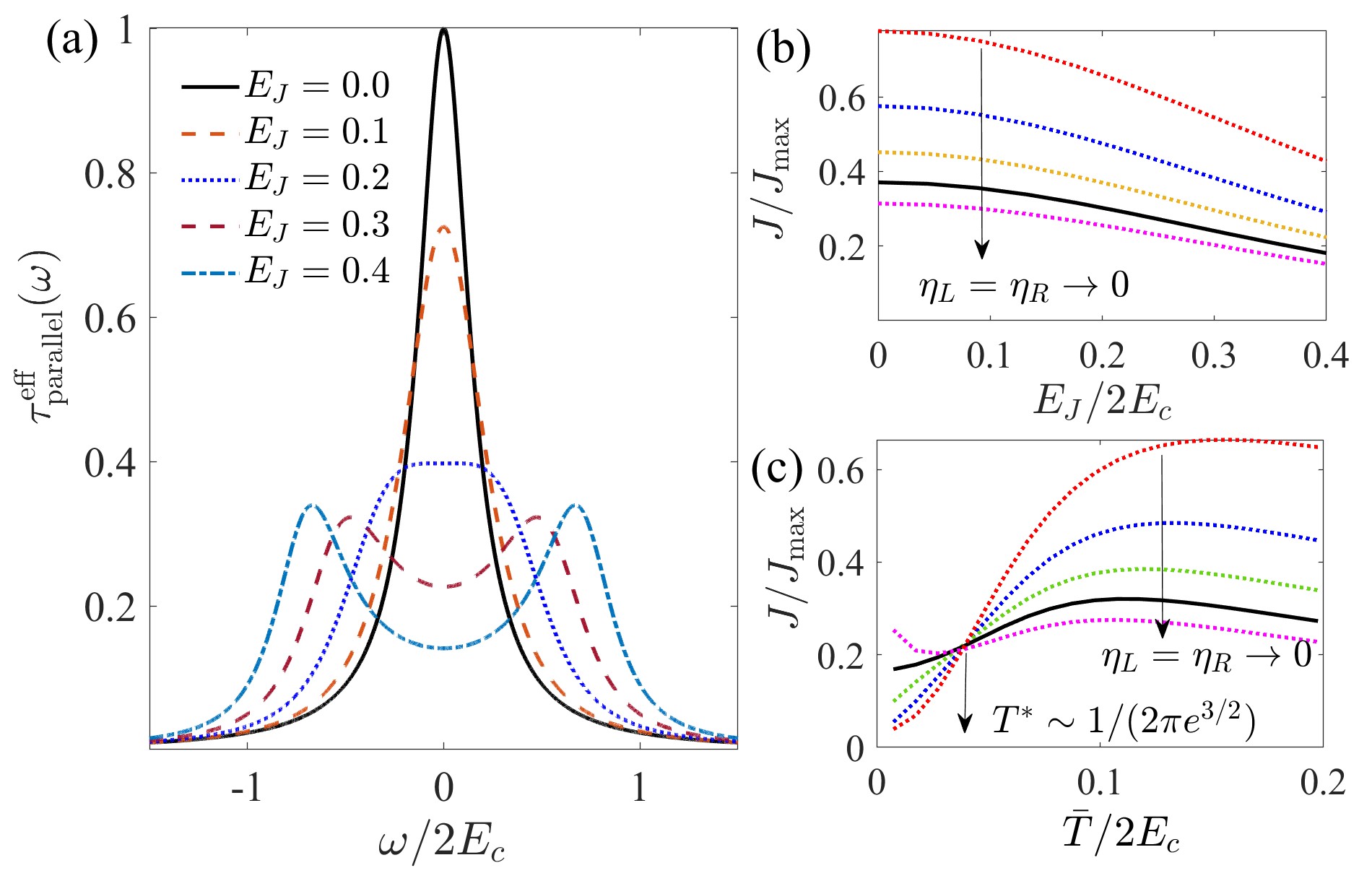}
\caption{(a) Effective heat transmission coefficient in the parallel configuration for increasing $E_J$ values (from 0 to 0.4) and $T_L=0.2$, $T_R=0.1$, $\omega_c= 2\times 10^{-3}$ (in units of $2E_c$) and $\eta_{L,R}=1/4\pi$. (b) Total heat current $J$, normalized to the maximum heat current for a ballistic channel $J_{\rm max}$, as a function of $E_J$. The full line corresponds to the parameters in (a), while the dashed lines illustrate its behavior with $\eta_L=\eta_R$ varying between $1/\pi$ to $1/5\pi$. (c) $J/J_{\rm max}$ for $E_J=0.2$ as a function of mean temperature $\bar{T}=(T_L+T_R)/2$ in the linear response regime for the same $\eta_L=\eta_R$ range.}
\label{fig3}
\end{figure}

Replacing these results into Eq.~\eqref{dressedGFs}, we obtain the results in Fig.~\ref{fig3}(a) for the heat transmission coefficient, which exhibits a strong dependence on $E_J$, in spite of the full suppression of the dc Josephson effect that occurs in the insulating phase \cite{ingold:1992} for this parameter range. As shown in Fig. \ref{fig3}(b), the total heat current in this configuration decreases with increasing $E_J$, and, as illustrated by the dashed lines in this panel, the sensitivity of the total heat current with $E_J$ exhibits a strong dependence on the lead resistances. 

One gets more insight by analyzing the low frequency renormalization of the heat transmission coefficient $\tau_{\rm parallel}(\omega)$. In fact, at low frequencies $\Sigma^{r(2)}(\omega) \sim -i \delta\eta \omega + \delta m \omega^2$, where $\delta \eta$ and $\delta m$ provide renormalization of the damping and mass model parameters. An analytical expression for the temperature and resistance scaling of these effective parameters is presented in \cite{SM}. At low temperatures ($T_{L,R} \ll \gamma$) and for $\eta_{L,R} \sim 1/2\pi$ we obtain
\begin{align}
\delta\eta&\sim \left(\frac{E_J}{\gamma}\right)^2 
\left(\frac{\eta_L T_L + \eta_R T_R}{m^2 \gamma^2}\right)^{-2}\prod_j \left(\frac{T_j}{\gamma}\right)^{\eta_j/\pi m^2 \gamma^2}\,,
\end{align}
where $\gamma = (\eta_L+\eta_R)/m$. 
Notice that for $T_L=T_R=T$ and $\alpha_L=\alpha_R=\alpha/2$, the SB transition at $\alpha=1$ is encoded in the scaling law $\sim T^{2/\alpha-2}$, which results for the renormalization of the damping parameter \cite{Aslangul1987,eckern_quantum_1987}.  

In terms of these renormalized parameters $\tau_{\rm parallel}(\omega)$ at low frequencies can be written as
\begin{equation}
\label{eq:taueffpar}
\tau^{\rm eff}_{\rm parallel}(\omega) \simeq \frac{4 \eta_L \eta_R}{\tilde{m}^2 \omega^2 + \tilde{\eta}^2} \frac{\tilde{\eta}}{\eta_L + \eta_R} \;,
\end{equation}
where $\tilde{\eta} = \eta_L + \eta_R + \delta \eta$ and $\tilde{m} = m - \delta m$. This expression accounts for the qualitative behavior of the heat transmission coefficient that can be observed in Fig. \ref{fig3}(a), i.e. a suppression and broadening of the $\omega \sim 0$ peak for increasing $E_J$. Notice that the peaks observed in the transmission coefficient for higher frequencies and finite $E_J$ (see Fig. \ref{fig3}) arise from the self-energy structure, which includes a mass renormalization that saturates for $\omega/2E_c \sim 1$.

On the other hand, lower temperatures are required to observe signatures of the SB transition in the heat transport properties. In Fig. \ref{fig3}(c) we show the heat current $J$ as a function of $\bar{T} = (T_L+T_R)/2$ in the linear regime ($\delta T = T_L-T_R \rightarrow 0$) for different values of $\eta_L=\eta_R$, ranging from $1/\pi$ to $1/5\pi$, i.e. across the transition which for the parallel case should occur at $\eta_{L,R} = 1/4\pi$. While for $\bar{T} \gtrsim \gamma$, $J$ decreases with increasing resistance due to the narrowing of the $\tau^{\rm eff}_{\rm parallel}(\omega)$ zero-frequency peak, the tendency is reversed at low temperatures ($\bar{T} \ll \gamma$) due to the divergence of $\delta \eta$ for $\alpha > 1$. The transition between these two opposite behaviors occurs around $\bar{T}= T^* \sim 1/(2\pi e^{3/2})$, where the damping renormalization is weakly dependent on the environment resistance \cite{SM}. This behavior is similar to the non-monotonous dependence of the mobility found for quantum Brownian motion in a periodic potential \cite{Fisher1985}.

\begin{figure}[t]
\includegraphics[width=\columnwidth]{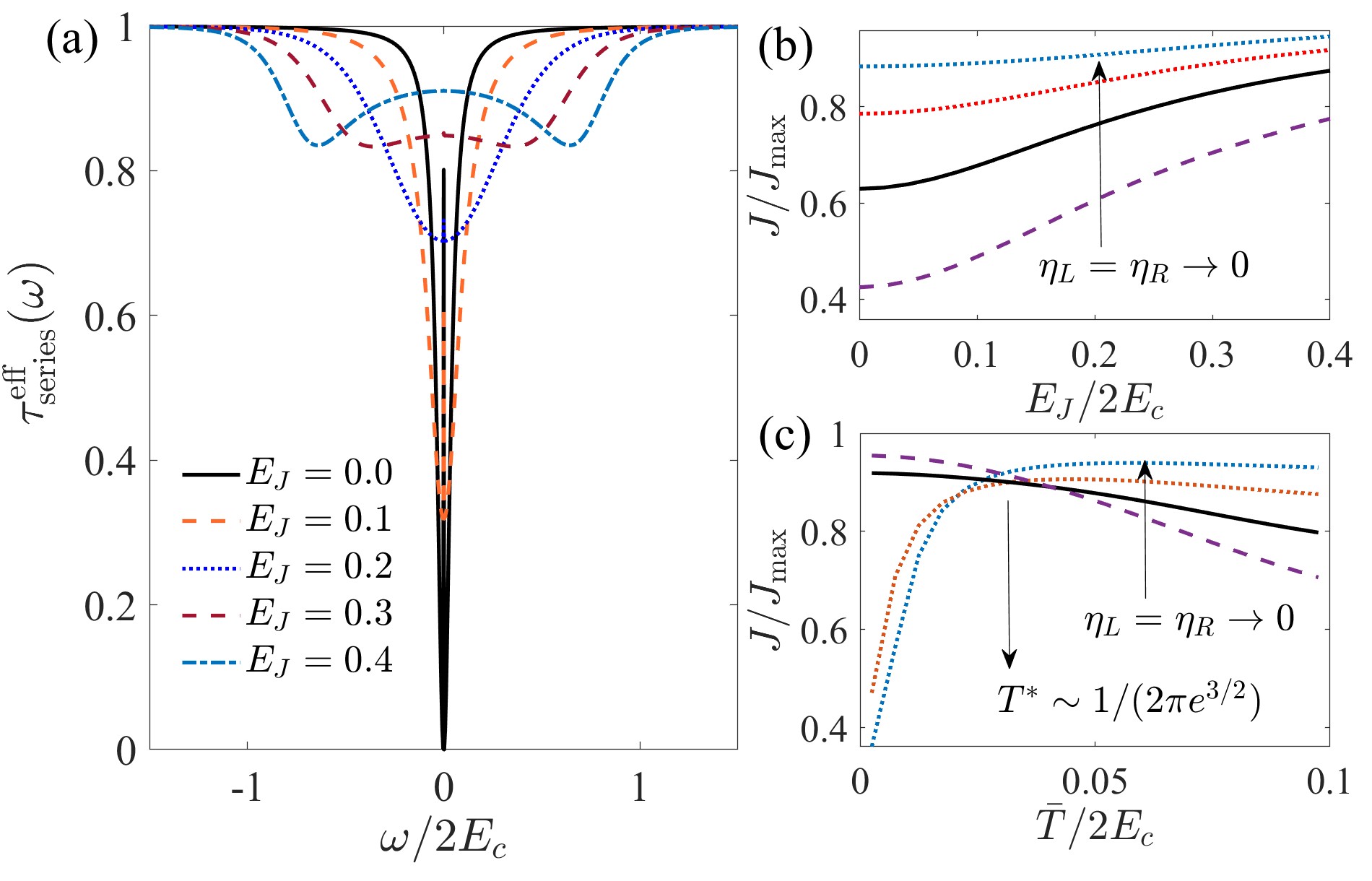}
\caption{Same as Fig. \ref{fig3} but for the series configuration. The parameters in (a) are the same as those in Fig. \ref{fig3}(a) while in (b) and (c) the full lines correspond to $\eta_L=\eta_R=1/\pi$, where the SB transition occurs for this configuration.}
\label{fig4}
\end{figure}

{\it Series configuration.---} In the actual setup in Ref.~\cite{subero_bolometric_2022} left and right leads are placed in a loop for the heat current measurements, which actually corresponds to a {\it series} connection, as depicted in Fig.~\ref{fig1}(b). One should describe this situation in terms of two phase variables $\varphi_1$ and $\varphi_2$, each one coupled to the left and right reservoirs respectively, while the charge $N$ through the junction is conjugate to $\varphi_1-\varphi_2$. Consequently, Eq.~\eqref{caldeira-leggett} should be modified by replacing the Josephson term by $-E_J \cos \left(\varphi_1-\varphi_2\right)$ and $\varphi$ by $\varphi_j$ in the coupling to the environmental modes.
Thus, the bare GFs adopt a matrix form in the $\varphi_{1,2}$ space and can be obtained from the following equations
\begin{eqnarray}
\left(\begin{array}{cc} m(\omega^2 - \omega_c^2)\pm i\omega\eta_L & -m(\omega^2 - \omega_c^2) \\
-m(\omega^2 -\omega_c^2) & m(\omega^2-\omega_c^2) \pm i\omega\eta_R \end{array} \right) \hat{D}^{r,a}_0 &=
& \hat{I}, \;
\end{eqnarray}
and $\hat{D}^K_0 = \hat{D}^r_0 \hat{d}^K_0 \hat{D}^a_0$, where
\[ \hat{d}^K_0 = -2i\omega \; \mbox{diag} \left[\eta_L \coth(\beta_L\omega/2); \eta_R \coth(\beta_R\omega/2) \right] \;.\]


To include the effect of finite $E_J$ in this configuration the corresponding self-energies in Eq. (\ref{sigK}) have to be expressed in terms of $\tilde{D}_0(t) = \mbox{Tr}\left[\hat{D}_0(t)(\hat{I}-\sigma_x)\right]$, where $\sigma_x$ is a Pauli matrix in $\varphi_{1,2}$ space. The resulting self-energies $\tilde{\Sigma}^{r,K}(\omega)$ exhibit a similar behavior as in the parallel configuration (see SM \cite{SM}).

On the other hand, the Dyson equations for the dressed GFs (\ref{dressedGFs}) acquire a matrix form in the $\varphi_{1,2}$ space and the corresponding self-energies are obtained from $\tilde{\Sigma}^{r,a,K}$ as $\hat{\Sigma}^{r,a,K} = \tilde{\Sigma}^{r,a,K}(\hat{I} - \sigma_x)$.

The heat current in the series connection cannot be expressed in the compact form (\ref{heat-MW}). It is, in contrast, necessary to reproduce the steps leading to Eq.~\eqref{heat-keldysh}, in order to derive the corresponding expression for the heat current in the series connection
\begin{eqnarray}
J_j &=& \int d\omega \omega^2 \eta_j \mbox{Im}\left[ D_{l_jl_j}^r(\omega)\left(1{+}2n_j(\omega)\right) - \frac{D_{l_jl_j}^K(\omega)}{2}\right], 
\label{heat-keldysh-series}
\end{eqnarray}
where $l_j=1,2$ for $j=L,R$ respectively. This expression can be decomposed~\cite{SM} as $J_j=(-1)^{l_j}J^{\rm el} + J^{\rm in}_j$, where $J^{\rm el} = \int d\omega \hbar \omega \tau^{\rm eff}_{\rm series}(\omega) \left[n_L(\omega) - n_R(\omega)\right]$ corresponds to the {\it elastic} contribution with
\begin{equation}
 \tau^{\rm eff}_{\rm series}(\omega) = 4\eta_L\eta_R \omega^2 |D^r_{12}(\omega)|^2 \;,
\end{equation}
while the {\it inelastic} contribution is given by
\begin{equation}
 J^{\rm in}_j = \int d\omega \omega^2 \eta_j \mbox{Im}\left[\hat{D}^r\left\{\left(\hat{\Sigma}^r{-}\hat{\Sigma}^a\right)(1{+}2n_j) - \hat{\Sigma}^K\right\}\hat{D}^a\right]_{l_jl_j} \;.
\end{equation}

In the non-interacting case (i.e. $E_J=0$) the inelastic term vanishes and the heat current can again be written in the usual Landauer form with a transmission coefficient at $\omega_c \rightarrow 0$
\begin{equation}
\tau^0_{\rm series}(\omega) = \frac{4 m^2 \omega^2 \eta_L \eta_R}{\left((\eta_L+\eta_R)m\omega\right)^2 + \left(\eta_L\eta_R\right)^2} \;,
\end{equation}
which exhibits a zero frequency dip and tends to $4\eta_L\eta_R/(\eta_L+\eta_R)^2$ at large frequencies, as illustrated in Fig. \ref{fig4}.

The behavior of the effective transmission coefficient for finite $E_J$ is illustrated in Fig. \ref{fig4}(a) for the same parameters as in Fig. \ref{fig3}(a). We observe that at finite $E_J$ the zero frequency dip is progressively suppressed. On the other hand, the inelastic term provides an additional contribution to the heat current \cite{SM}.
Thus, in contrast to the parallel configuration, there is a slight increase of the heat current with $E_J$, which is confirmed by the integrated results in Fig. \ref{fig4}(b). As can be observed, this increase is more pronounced as the leads resistances approach $R_Q$. This behavior is in qualitative agreement with the experimental results in Ref.~\cite{subero_bolometric_2022}. On the other hand, the temperature dependence of $J/J_{\rm max}$ in the linear regime, illustrated in Fig.~\ref{fig4}(c), exhibits a crossover at $T^*$ as in the parallel case but with an opposite tendency with environmental resistance: it increases with increasing resistance (decreasing $\eta_{L,R}$) for $\hat{T} > T^*$ and the opposite for $\bar{T} < T^*$. At the critical resistance ($\eta_L=\eta_R=1/\pi$) the temperature variation of the heat conductance is minimal, which provides an alternative signature of the transition. Notice that the experimental results of Ref. \cite{subero_bolometric_2022} exhibit a drop at the lower temperatures consistent with an insulating behavior (see additional figure in the SM \cite{SM}).


{\it Rectification properties.---} Another interesting property of a JJ in a resistive environment is its possible behavior as a heat rectifier. The JJ anharmonicity has been used to define thermal diodes based on weakly coupled qubits~\cite{segal_spin_2005}, 
see also Ref.~\cite{ojanen_mesoscopic_2008}.
Though heat rectifiers using JJs have been measured in the absence of environmental effects (but additionally coupled to a phonon bath) \cite{martinez-perez2013} or in superconducting qubits~\cite{senior_heat_2020,upadhyay_microwave_2024} (see also Ref.~\cite{liliana_review} for a review), their rectifying properties in the insulating side of the SB transition have not been considered. For simplicity we consider here only the parallel configuration with an asymmetry provided by $R_L \neq R_R$. We can then define a ``forward" ($J^+$) and a ``reverse" ($J^-$) heat current given by
\begin{equation}
    J^{\pm} = \int \frac{d\omega}{\pi} \hbar \omega \tau^{\rm eff (\pm)}_{\rm parallel}(\omega) \left[n^{\pm}_L(\omega) - n^{\pm}_R(\omega)\right] \;,
    \label{heat-asymmetry}
\end{equation}
which correspond to positive and negative temperature bias $\Delta T = T_L-T_R$. As is customary, we also define a heat rectification ratio ${\cal R} = |J^+/J^-|$ which, as can be seen from Eq. (\ref{heat-asymmetry}), deviate from unity provided $\delta \tau^{\rm eff} \equiv \tau^{\rm eff(+)}_{\rm parallel} - \tau^{\rm eff(-)}_{\rm parallel} \neq 0$. 

\begin{figure}[t]
\includegraphics[width=1\columnwidth]{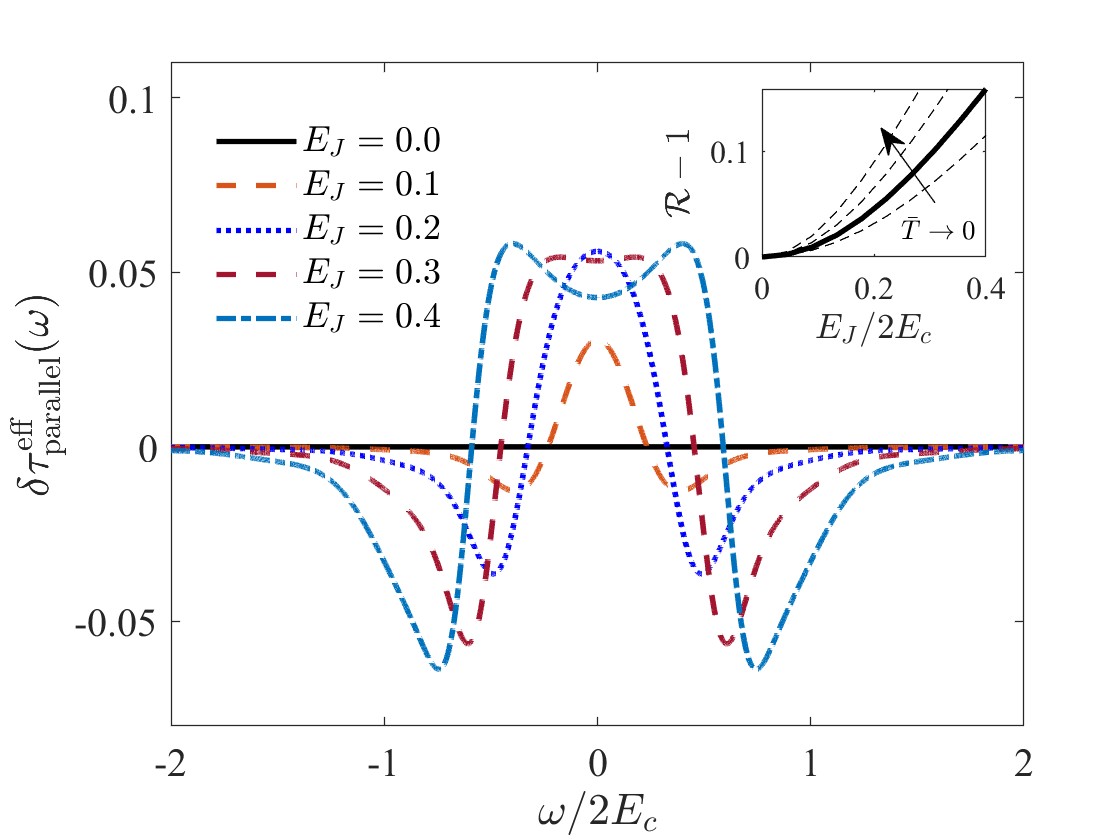}
\caption{Asymmetry in the heat transmission coefficient $\delta \tau^{\rm eff}(\omega) = \tau^{\rm eff(+)}_{\rm parallel}(\omega)-\tau^{\rm eff(-)}_{\rm parallel}(\omega)$ for the parallel configuration with $\eta_L=1/\pi$, $\eta_R=1/4\pi$, $\Delta T=0.1$, ${\bar T} = (T_L+T_R)/2=0.15$, and increasing values of $E_J$ from 0 to 0.4. The corresponding heat rectification ratio ${\cal R} = |J^+/J^-|$ as a function of $E_J$ is shown as a full line in the inset. The dashed lines illustrate its variation with $\bar{T}$ from 0.175 to 0.1.}
\label{fig5}
\end{figure}

In Fig. \ref{fig5} we show the behavior of $\delta \tau^{\rm eff}(\omega)$ for increasing values of $E_J$ in an asymmetric configuration with $\alpha_L = 2$ and $\alpha_R = 0.5$. We observe that, for finite $E_J$, heat transport in the forward direction is favored at low frequencies, while the opposite occurs at higher frequencies. There is not, however, a compensation in the total heat current and thus one obtains that ${\cal R} > 1$ as illustrated in the inset of Fig. \ref{fig5}. Rectification ratios of the order of $10\%$ or larger are obtained for the set of parameters in this figure. These ratios can increase further by decreasing the mean temperature $\bar{T}=(T_L+T_R)/2$ 
(see inset in Fig. \ref{fig5}).       

Further insight on these properties can be obtained from the analytical approximation to the system self-energies. As shown in \cite{SM} the rectification ratio in the parallel configuration can be expressed as
\beq
{\cal R}-1\sim(mE_J)^2\frac{\eta_R-\eta_L}{\eta_L+\eta_R}\left(\frac{2}{\alpha}-2\right)\bar{T}^{\frac{2}{\alpha}-2}\frac{\delta T}{\bar{T}}
\label{rect-anal}
\eeq
in the limit $T_{L,R} \ll \gamma$ and $\alpha \gtrsim 1$ upon a small temperature difference $\delta T\ll \bar{T}$.
This expression accounts for the main features of the numerical results, i.e. the quadratic increase with $E_J$ and the increase of ${\cal R}$ as the mean temperature is decreased. Eq. (\ref{rect-anal}) also indicates that, as far as $\alpha > 1$, the heat current is larger when the colder reservoir corresponds to the one with larger resistance. This behavior is similar to the case of a two-level system with a nonseparable coupling to the heat baths~\cite{segal_spin_2005,rect}. Eq. (\ref{rect-anal}) also indicates that the SB transition could be detected by the change in sign of ${\cal R} -1$ at $\alpha = 1$.

{\it Conclusions.---} In summary, we have presented non-equilibrium Green functions calculations for photonic heat transport through a Josephson junction in a resistive environment. Our results are in qualitative agreement with the experimental ones from \cite{subero_bolometric_2022} and suggest how signatures of the SB transition can be detected in heat transport measurements at sufficiently low temperatures \cite{note-alf}. We further demonstrate rectification properties for this device that can be tested 
in future experiments.

{\it Note added:} While completing this manuscript we become aware of a related study \cite{Yamamoto2024} for the same system in the scaling regime, yielding results which are complementary to the ones in the present work.   

The authors wish to thank M. Houzet and P. Joyez for enlightening discussions. ALY also acknowledges Prof. U. Weiss for useful comments on the validity of the perturbative approach adopted in this work. We acknowledge funding from the Spanish Ministerio de Ciencia e Innovaci\'on via grants No. No.~PID2020-117671GB-I00, PID2019-110125GB-I00 and PID2022-142911NB-I00, and through the ``Mar\'{i}a de Maeztu'' Programme for Units of Excellence in R{\&}D CEX2023-001316-M. This work was partially funded by the Research Council of Finland Centre of Excellence program grant 336810 and grant 349601 (THEPOW).

\bibliography{biblio.bib}

\begin{thebibliography}{37}
\bibitem [{rec()}]{rect}%
  \BibitemOpen
  \href@noop {} {}\bibinfo {note} {Note that the temperature behaviour of the rectification is strongly system and parameter dependent.}\BibitemShut {Stop}%
\end{thebibliography}

\setcounter{equation}{0}
\renewcommand{\theequation}{S\,\arabic{equation}}

\setcounter{figure}{0}
\renewcommand{\thefigure}{S\,\arabic{figure}}

\onecolumngrid
\vspace{\columnsep}
\section*{Supplementary information for ``Photonic heat transport through a Josephson junction in a resistive environment"}
\vspace{\columnsep}
\twocolumngrid

\section{Details on numerical calculations}

\subsection{Self-energies}

From Eq. (8) in the main text, the zero order GFs in time representation are given by
\begin{eqnarray}
D_0^r(t) &=& -\frac{\theta(t)}{m\omega_1} e^{-\gamma t/2} \sinh\left(\omega_1 t\right) \nonumber \\
D_0^K(t) &=& S_1(t) + S_2(t)
\label{analytical-propagators}
\end{eqnarray}
where $\gamma = (\eta_L+\eta_R)/m$, $\omega_1 = \sqrt{\gamma^2/4 - \omega_c^2}$ and
\begin{eqnarray}
S_1(t) &=& -\sum_{\mu=L,R}\frac{\eta_{\mu}}{2m\omega_1\gamma}\left(\coth\left(\frac{\beta_{\mu} \lambda_1}{2}\right) e^{i\lambda_1|t|}- \coth\left(\frac{\beta_{\mu} \lambda_2}{2}\right) e^{i\lambda_2|t|}\right) \nonumber\\
S_2(t) &=& 4i \sum_{\mu=L,R} \frac{\eta_{\mu}}{m\beta_{\mu}} \sum_{n>0} \frac{\nu^{\mu}_n e^{-\nu^{\mu}_n |t|}}{\left(\omega_c^2 + (\nu^{\mu}_n)^2\right)^2 - \gamma^2(\nu^{\mu}_n)^2} \;,
\end{eqnarray}
where $\lambda_{1,2} = i (\gamma/2 \pm \omega_1)$, and $\nu_n^{\nu} = 2n/\beta_{\nu}$ are the Matsubara frequencies. 

By introducing these analytical expressions for the propagators in time representation into Eqs. (12) in the main text and performing their Fourier transform numerically, we obtain the results shown in Fig. 2 for the different self-energies in frequency space. For comparison we show the corresponding results for the series configuration. 

\begin{figure}[t]
\includegraphics[width=1\columnwidth]{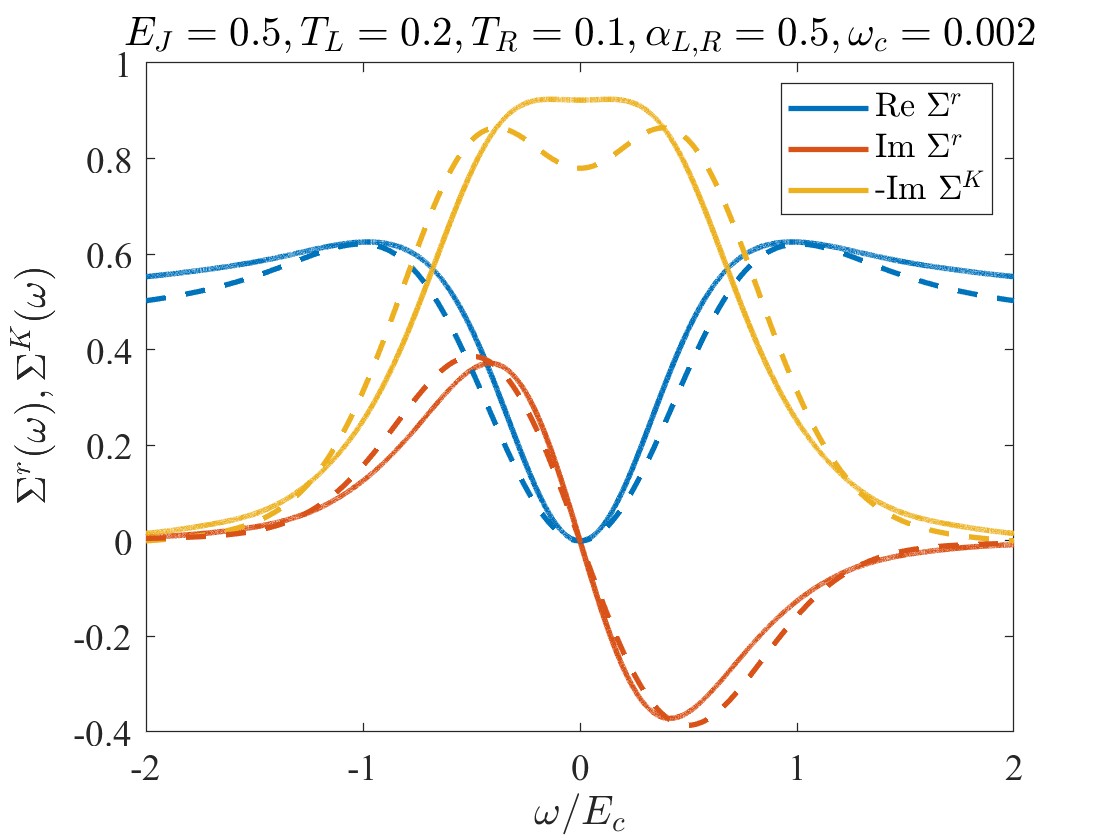}
\caption{Self-energies in frequency representation obtained numerically from Eqs. (12) in the main text. Energies in units of $E_c$. The corresponding parameters are indicated in the figure title. The dashed lines correspond to the self-energies in the series configuration for the same set of parameters.}
\label{fig2}
\end{figure}

\begin{figure}[h!]
\includegraphics[width=1\columnwidth]{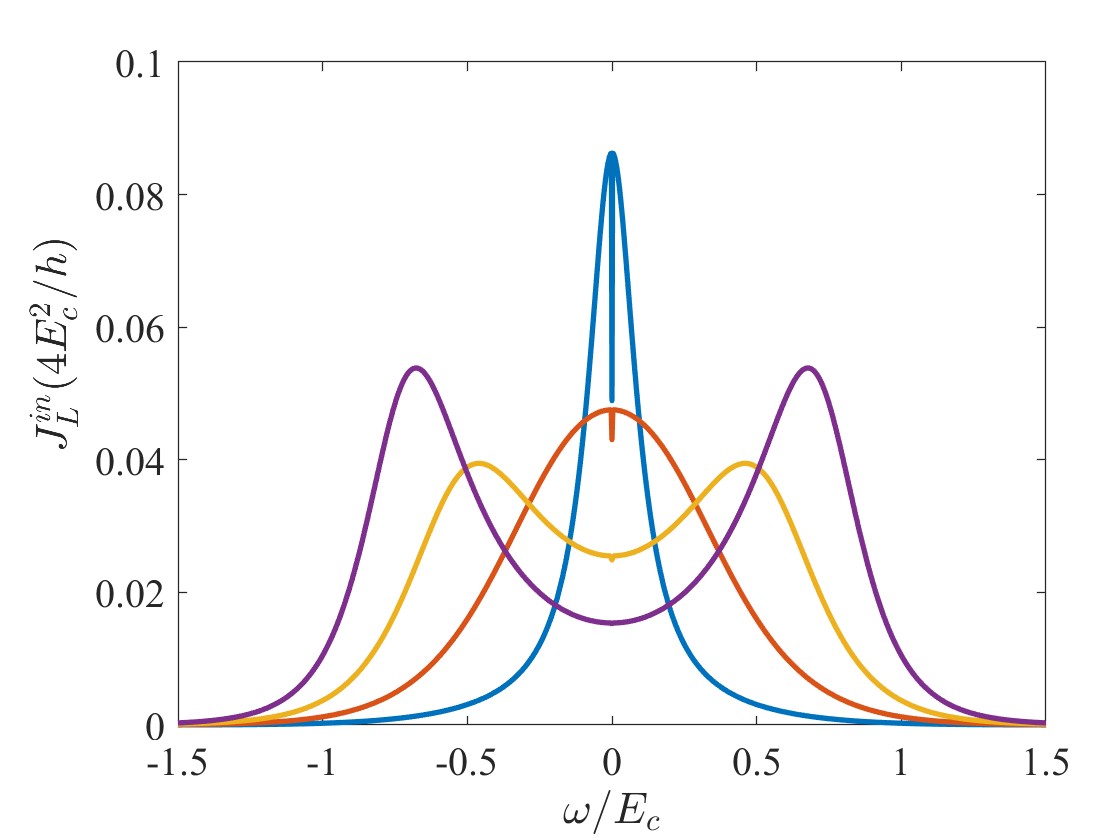}
\caption{Inelastic contribution to the heat current in the series configuration for the same parameters as in Fig. 3(a) in the main text.}
\label{inelastic}
\end{figure}

In Fig. \ref{inelastic} we show the inelastic contribution to the heat current in the series configurations for the same parameters as in Fig. 3(a) in the main text. As can be observed, this contribution is positive so that it tends to increase the total heat current when added to the elastic contribution.



\section{Analytical approximations}

Here we give details on the approximations used to obtain the analytic behaviour of the self-energies $\Sigma^{r(2)}(\omega)$ and $\Sigma^{K(2)}(\omega)$ in the parallel configuration. 

\subsection{Retarded self-energy}

\begin{figure}[t]
\includegraphics[width=\linewidth]{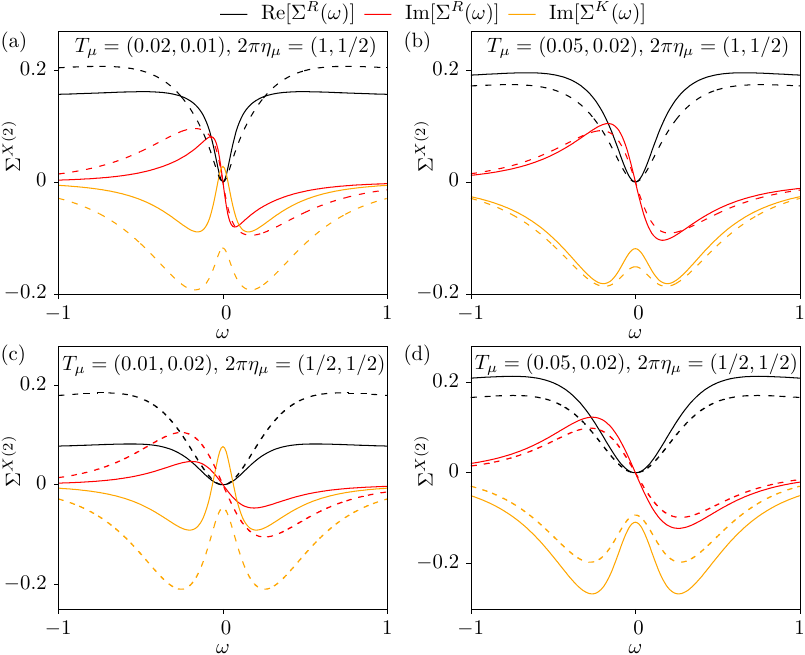}
\caption{Self-energies for different temperatures $T_\mu=(T_L,T_R)$ and damping parameters $\eta_\mu=(\eta_L,\eta_R)$. (a) and (b): asymmetric case with $\eta_L=1/2\pi$ and $\eta_R=1/4\pi$. (c) and (d): symmetric case with $\eta_L=\eta_R=1/4\pi$.
Dashed lines show the numerical results. 
In all panels, $E_J=0.2$, $m=1$. 
}\label{fig:selfen}
\end{figure}

We start with the time-dependent retarded selfenergy as expressed in Eq.~(12) in the main text:
\beq
\label{eq:SigRt}
\Sigma^{r(2)}(t) = E_J^2 \sin \frac{D_0^r(t)}{2} \exp \left[\frac{i}{2}\left(D_0^K(t) -D^K_0(0)\right)\right] - B \delta(t),
\eeq
of which we want to obtain the Fourier transform. 
For this, we consider the first term:
\beq
\tilde\Sigma(t)=E_J^2 \sin \frac{D_0^r(t)}{2} \exp \left[\frac{i}{2}\left(D_0^K(t) -D^K_0(0)\right)\right].
\eeq
The second term, proportional to $B$ will be cancelled out with $\tilde\Sigma(0)$.
We can simplify it by writing $D_0^r(t)=D_0^a(-t)=-\theta(t)(1-\exp[-\gamma t])/m\gamma$, and using the approximation 
\beq
\label{eq:eckernappr}
{\rm Re}\left[\frac{i}{2}(D_0^K(t){-}D_0^K(0))\right]\approx-\sum_\mu\frac{\eta_\mu}{\pi m^2\gamma^2}\ln\left[\frac{\gamma\beta_\mu}{\pi}\sinh\left(\frac{\pi|t|}{\beta_\mu}\right)\right]
\eeq
for $|\gamma t|\gg1$ and $\kBT\ll\gamma$~\cite{eckern_quantum_1987}. Doing so, we get:
\beq
\label{eq:SigRtapp}
\tilde\Sigma(t)=\theta(t)E_J^2 \sin\left(\frac{e^{-\gamma t}-1}{2m\gamma}\right)\prod_\mu\left[\frac{\gamma\beta_\mu}{\pi}\sinh\left(\frac{\pi|t|}{\beta_\mu}\right)\right]^{-\eta_\mu/\pi m^2\gamma^2},
\eeq
where $\eta_\mu=R_Q/2\pi R_\mu$, $m=\hbar/E_c$ and $\gamma=\sum_\mu\eta_\mu/m$, as defined in the main text. 
We need to solve its Fourier transform:
\beq
\tilde\Sigma(\omega)={\cal A}\gamma\int_0^\infty dt\sin\left(\frac{e^{-\gamma t}{-}1}{2m\gamma}\right)\prod_\mu\left[2\sinh\left(\frac{\pi|t|}{\beta_\mu}\right)\right]^{-\frac{\eta_\mu}{\pi m^2\gamma^2}}e^{i\omega t},
\eeq
where we have defined the prefactor
\beq
{\cal A}\equiv \frac{E_J^2}{\gamma}\prod_\mu\left(\frac{\gamma\beta_\mu}{2\pi}\right)^{-\eta_\mu/\pi m^2\gamma^2}.
\eeq
To reach the low frequency limit $T_j \gg \omega$, we approximate $\sinh(\pi|t|/\beta_{\mu})\approx e^{\pi|t|/\beta_{\mu}}/2$. Then, performing the change of variables $u=e^{-\gamma t}$ we get:
\beq
\label{eq:Sigsimpl}
\tilde\Sigma(\omega)={\cal A}\int_0^1 du\sin\left(\frac{u{-}1}{2m\gamma}\right)u^{c-1},
\eeq
with 
\beq
c\equiv \frac{1}{m^2\gamma^3}\left(\frac{\eta_L}{\beta_L}+\frac{\eta_R}{\beta_R}\right)-\frac{i\omega}{\gamma}.
\eeq
The integral in Eq.~\eqref{eq:Sigsimpl} has an analytical solution if Re$(c)>0$:
\beq
\int_0^1 du\sin[a(u{-}1)]u^{c-1}=g_1(a,c)+g_2(a,c),
\eeq
with
\begin{gather}
\label{eq:g}
\begin{aligned}
g_1(a,c)&=\frac{\cos(a)}{(1{+}c)/a}~{_1F_2}\left(\frac{1{+}c}{2};\frac{3}{2},\frac{3{+}c}{2};-\frac{a^2}{4}\right)\\
g_2(a,c)&=-\frac{\sin(a)}{c}~{_1F_2\left(\frac{c}{2};\frac{1}{2},\frac{2{+}c}{2};-\frac{a^2}{4}\right)},
\end{aligned}
\end{gather}
and where $_1F_2(a_1;b_1,b_2;z)$ is a ``generalized hypergeometric function". Formally, this then gives an analytical solution:  
\begin{align}
\label{eq:tildSanal}
\tilde\Sigma(\omega)={\cal A}\left[g_1\left(\frac{1}{2m\gamma},c\right)+g_2\left(\frac{1}{2m\gamma},c\right)\right],
\end{align}
but it is not very informative until we plot its real and imaginary parts, which is done in Fig.~\ref{fig:selfen}.

\begin{figure}[t]
\includegraphics[width=\linewidth]{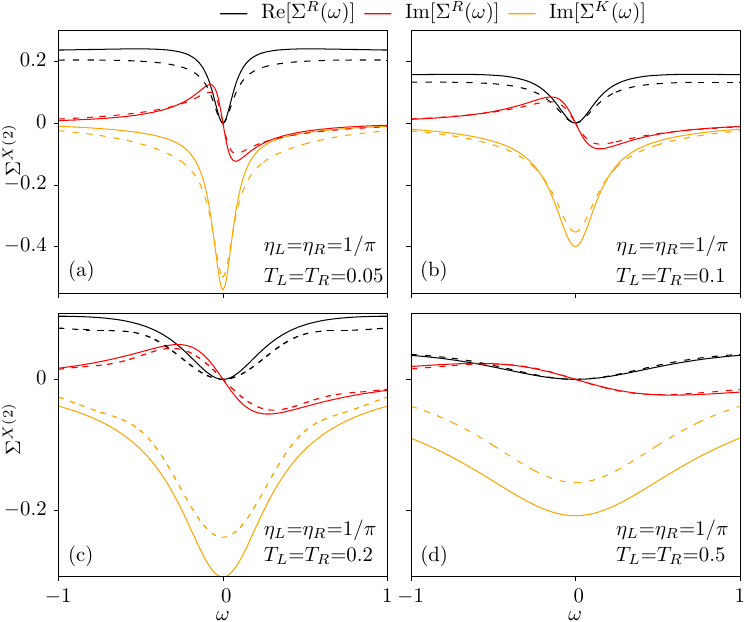}
\caption{
Comparison of numeric and analytic selfenergies in the low impedance regime, $R_\mu=0.5R_Q$, for increasing temperatures. Other parameters are as in Fig.~\ref{fig:selfen}.
}\label{fig:selfenlowR}
\end{figure}

In a linear expansion, the constant term is real, giving the value of $B$ in Eq.~\eqref{eq:SigRt}, and the linear term is pure imaginary, so the linear expansion simplifies to:
\beq
\label{eq:tildSanal_lin}
\tilde\Sigma(\omega)=i{\cal A}{\rm Im}\left[\left.(g'_1+g'_2)\right|_{\omega=0}\right]\omega.
\eeq 

\subsection{Keldysh self-energy}

Now we consider the time-dependent Keldysh self-energy in Eq.~(12) in the main text:
\begin{gather}
\begin{aligned}
\label{eq:SigKt}
\Sigma^{K(2)}(t) = \frac{E_J^2}{i} \cos\left(\frac{D_0^r(t)}{2}\right)\cos\left(\frac{D_0^a(t)}{2}\right)\exp \left[\frac{i}{2}\left(D_0^K(t){-}D^K_0(0)\right)\right].
\end{aligned}
\end{gather}
We separate the time integral in the Fourier transform in two terms, with $D_0^r(t<0)=0$ and $D_0^a(t>0)=0$:
\begin{align}
\Sigma^{K(2)}(\omega) =& -i{\cal A}\gamma\left\{\int_{-\infty}^0dt \cos\left(\frac{e^{\gamma t}{-}1}{2m\gamma}\right)\prod_\mu\left[2\sinh\left(\frac{-\pi t}{\beta_\mu}\right)\right]^\frac{-\eta_\mu}{\pi m^2\gamma^2}e^{i\omega t}\right.\nonumber\\
&\left.+\int_{0}^{\infty}dt \cos\left(\frac{e^{-\gamma t}{-}1}{2m\gamma}\right)\prod_\mu\left[2\sinh\left(\frac{\pi t}{\beta_\mu}\right)\right]^\frac{-\eta_\mu}{\pi m^2\gamma^2}e^{i\omega t}\right\},
\label{eq:SigKsep}
\end{align}
where we have again used the approximation in Eq.~\eqref{eq:eckernappr}~\cite{eckern_quantum_1987}.
Doing a change $t\rightarrow-t$ in the first integral of Eq.~\eqref{eq:SigKsep}, we can write
\beq
\Sigma^{K(2)}(t) = -i{\cal A}\gamma\left({\cal I}^++{\cal I}^-\right),
\eeq
with
\beq
{\cal I}^\pm=\int_{0}^{\infty}dt \cos\left(\frac{e^{-\gamma t}{-}1}{2m\gamma}\right)\prod_\mu\left[2\sinh\left(\frac{\pi t}{\beta_\mu}\right)\right]^\frac{-\eta_\mu}{\pi m^2\gamma^2}e^{\pm i\omega t}.
\eeq
Following the same steps as for the retarded case ($u=e^{-\gamma t}$), we arrive to:
\begin{align}
\label{eq:SKappr}
\Sigma^{K(2)}=-i{\cal A}&\int_0^1 du\cos\left(\frac{u{-}1}{2m\gamma}\right)\left(u^{c-1}+u^{c^*-1}\right)
\end{align}
In this case, we can use the integral:
\beq
\int_0^1 du\cos[a(u{-}1)]u^{c-1}=g_3(a,c)+g_4(a,c),
\eeq
with 
\begin{gather}
\label{eq:g34}
\begin{aligned}
g_3(a,c)&=\frac{\cos(a)}{c}~{_1F_2\left(\frac{c}{2};\frac{1}{2},\frac{2{+}c}{2};-\frac{a^2}{4}\right)}\\
g_4(a,c)&=a\frac{\sin(a)}{1{+}c}~{_1F_2\left(\frac{1{+}c}{2};\frac{3}{2},\frac{3{+}c}{2};-\frac{a^2}{4}\right)},
\end{aligned}
\end{gather}
provided Re$(c)>0$.
We note that $g_3(1/2m\gamma,c^*)=g_3^*(1/2m\gamma,c)$ and $g_4(1/2m\gamma,c^*)=g_4^*(1/2m\gamma,c)$.
Hence, the Keldysh self-energy is purely imaginary:
\begin{align}
\Sigma^{K(2)}=-i2{\cal A}{\rm Re}\left[g_3\left(\frac{1}{2m\gamma},c\right)+g_4\left(\frac{1}{2m\gamma},c\right)\right].
\end{align}
The result is plotted in orange lines in Fig.~\ref{fig:selfen}.



\subsection{Frequency expansion}

We can expand on the integrand of Eqs.~\eqref{eq:Sigsimpl} and \eqref{eq:SKappr} for low frequencies, getting:
\beq
\Sigma^{\alpha(2)}\approx \Sigma_0^{\alpha(2)}+\omega\Sigma_1^{\alpha(2)}+\omega^2\Sigma_2^{\alpha(2)},
\eeq
with 
\begin{align}
\label{eq:sig1R}
\Sigma_k^{r(2)}&=-\frac{1}{k!}\left(\frac{i}{\gamma}\right)^k{\cal A}{\cal I}_k(c_0)\\
\Sigma_0^{K(2)}&=-2i{\cal A}\int_0^1 du\cos\left(\frac{u{-}1}{2m\gamma}\right)u^{c_0-1}\\
\Sigma_1^{K(2)}&=0\\
\Sigma_2^{K(2)}&=\frac{i}{\gamma^2}{\cal A}\int_0^1 du\cos\left(\frac{u{-}1}{2m\gamma}\right)u^{c_0-1}\ln^2u,
\end{align}
where
\beq
c_0\equiv c(\omega=0)= \frac{1}{m^2\gamma^3}\left(\frac{\eta_L}{\beta_L}+\frac{\eta_R}{\beta_R}\right)
\eeq
and
\beq
{\cal I}_k(c_0)=\int_0^1 du\sin\left(\frac{u{-}1}{2m\gamma}\right)u^{c_0-1}\ln^k u.
\eeq
These integrals again have analytical solutions in terms of hypergeometric functions of higher order. For this, we define:
\beq
{\cal G}_{pqs}^{(n)}={}_nF_{n+1}\left(\frac{p+c_0}{2}1_n;\frac{q}{2},\frac{s+c_0}{2}1_n;-\frac{1}{16m^2\gamma^2}\right),
\eeq
where $1_n$ is a list of $n$ units.With these, we write:
\begin{align}
\label{eq:Ik}
{\cal I}_0(c_0)&=\frac{\cos\left(\frac{1}{2m\gamma}\right)}{2m\gamma(1{+}c_0)}{\cal G}_{1,3,3}^{(1)}-\frac{\sin\left(\frac{1}{2m\gamma}\right)}{c_0}{\cal G}_{0,1,2}^{(1)}\nonumber\\
{\cal I}_1(c_0)&=-\frac{2m\gamma(1{+}2c_0)\cos\left(\frac{1}{2m\gamma}\right)\sin\left(\frac{1}{2m\gamma}\right)}{c_0^2(1{+}c_0)^2}\nonumber\\
&+\frac{\cos\left(\frac{1}{2m\gamma}\right)}{12m^2\gamma^2(1{+}c_0)(3{+}c_0)}\left[\frac{{\cal G}_{3,5,5}^{(1)}}{1{+}c_0}+\frac{{\cal G}_{3,5,5}^{(2)}}{3{+}c_0}\right]\\
&-\frac{\sin\left(\frac{1}{2m\gamma}\right)}{2m\gamma c_0^2(2{+}c_0)^2}\left[(2{+}c_0){\cal G}_{2,3,4}^{(1)}+c_0{\cal G}_{2,3,4}^{(2)}\right]\nonumber\\
{\cal I}_2(c_0)&=2\left[\frac{\cos\left(\frac{1}{2m\gamma}\right)}{2m\gamma(1{+}c_0)^3}{\cal G}_{2,3,3}^{(3)}-\frac{\sin\left(\frac{1}{2m\gamma}\right)}{c_0^3}{\cal G}_{0,1,2}^{(3)}\right].\nonumber
\end{align}
For the Keldysh self-energy, it is sufficient to look at the value at the origin, $\Sigma^{K(2)}(\omega)=\Sigma_0^{K(2)}+{\cal O}(\omega^2)$:
\begin{align}
\label{eq:SigKcoeff}
\Sigma_0^{K(2)}=-2i{\cal A}\left[\frac{\cos\left(\frac{1}{2m\gamma}\right)}{c_0}{\cal G}_{0,1,2}^{(1)}+\frac{\sin\left(\frac{1}{2m\gamma}\right)}{2m\gamma(1{+}c_0)}{\cal G}_{1,3,3}^{(1)}\right].
\end{align}

\subsubsection{Low temperature scaling}


The linear coefficient of the retarded self-energy corresponds to the damping parameter correction, $\delta \eta$, defined in the main text. Taking the low temperature limit in the above expressions it is straightforward to show that for $T_{L,R} \ll \gamma$ and $m\gamma \sim 1/(2\pi)$

\beq
\delta \eta \sim {\cal A}\frac{\gamma}{c_0^2} \equiv \left(\frac{E_J}{\gamma}\right)^2  \frac{\left(\frac{T_L}{\gamma}\right)^{\frac{\eta_L}{\pi m^2 \gamma^2}}\left(\frac{T_R}{\gamma}\right)^{\frac{\eta_R}{\pi m^2 \gamma^2}}}{\left(\frac{\eta_L T_L + \eta_R T_R}{m^2 \gamma^2}\right)^2} \;.
\label{scaling}
\eeq

In the limit $T_L=T_R=T$ the scaling reduces to $\delta \eta \sim T^{2/\alpha-2}$, where $\alpha = 2\pi(\eta_L+\eta_R)$ is the inverse of the total resistance in units of $R_Q$, which signals the SB transition at $\alpha=1$ \cite{eckern_quantum_1987}. 

Based on this scaling it is possible to determine the transition point $T^*$ for heat transport indicated in Figs. 2(c) and 3(c) of the main text. For that purpose remind that at low frequency the heat transmission coefficient scales as $\eta/\bar{\eta}$ for a symmetric case with $\eta_L=\eta_R=\eta$ (see Eq. (14) in main text). Thus, in order to get the same heat current for two cases with different resistances at the same mean temperature we need
\begin{eqnarray}
\frac{\bar{\eta}_1}{\eta_1} = \frac{\bar{\eta}_2}{\eta_2} \longrightarrow \frac{T^{2/\alpha_1-2}}{\gamma_1^{2/\alpha_1+1}} = \frac{T^{2/\alpha_2-2}}{\gamma_2^{2/\alpha_2+1}} \;.
\end{eqnarray}

Setting $\alpha_1=1$ in the limit $\alpha_2 \rightarrow 1$ we obtain $T=T^*=1/(2\pi e^{3/2}) \simeq 0.036$, in agreement with the numerical results in Figs. 2(c) and 3(c) of the main text. 


\section{Rectification}

\begin{figure}[t]
\includegraphics[width=.5\linewidth]{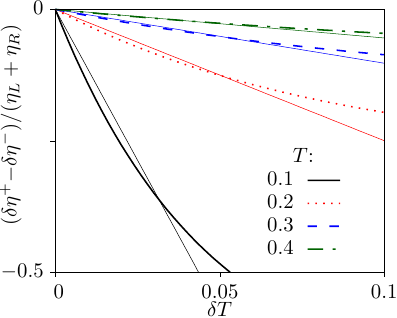}
\caption{Rectification effect as a function of the temperature difference using the approximation of Eq.~\eqref{eq:rectnum}, for different temperatures. Thin solid lines correspond to the linear expansion of Eq.~\eqref{eq:scaleta}. Parameters: $E_J=0.2$, $m=1$, $\eta_L=1/\pi$, $\eta_R=1/4\pi$.}
\label{fig:detaFB}
\end{figure}

The term $\Sigma_1^{r(2)}$ modifies the damping parameter $\Sigma_1^{r(2)}=-i\delta\eta$, which affects the transmission probability. If the resistances are different, the response is different depending on their temperature. Linearizing Eq.~(14) in the main text at $\omega=0$:
\beq
\tau_{\rm eff}^\pm\approx\frac{4\eta_L\eta_R}{(\eta_L+\eta_R)^2}\left(1-\frac{\delta\eta^\pm}{\eta_L+\eta_R}\right).
\eeq
A rectification effect appears when $\delta\tau_{\rm parallel}^{\rm eff}=\tau_{\rm eff}^+-\tau_{\rm eff}^-\neq0$:
\beq
\delta\tau_{\rm parallel}^{\rm eff}\approx\frac{4\eta_L\eta_R}{(\eta_L+\eta_R)^3}\left(\delta\eta^--\delta\eta^+\right).
\label{eq:rectnum}
\eeq
We hence need
\beq
\delta\eta^--\delta\eta^+=i\left[\Sigma_{1,L}^{r(2)}-\Sigma_{1,R}^{r(2)}\right]
\eeq
to be finite, where $\Sigma_{1,\mu}^{r(2)}$ is defined as the linear coefficient $\Sigma_{1}^{r(2)}$ when terminal $\mu$ is at temperature $T+\delta T$ and the other one is at $T$. They are obtained using Eqs.~\eqref{eq:sig1R} and \eqref{eq:Ik}. The result is plotted in Fig.~\ref{fig:detaFB} as a function of the temperature difference $\delta T$.

We gain further insight by assuming a small temperature difference $\delta T$, so we can linearize use the scaling of Eq.~\eqref{scaling}, getting 
\beq
\label{eq:scaleta}
\delta\eta^--\delta\eta^+\approx
(mE_J)^2\frac{\eta_R-\eta_L}{\eta_L+\eta_R}\left(\frac{2}{\alpha}-2\right)T^{\frac{2}{\alpha}-2}\frac{\delta T}{T},
\eeq
see Fig.~\ref{fig:detaFB}.
This expression gives the scaling of the rectification coefficient:
\beq
{\cal R}-1\sim(mE_J)^2\frac{\eta_R-\eta_L}{\eta_L+\eta_R}\left(\frac{2}{\alpha}-2\right)T^{\frac{2}{\alpha}-2}\frac{\delta T}{T}
\eeq
given in the main text.

\subsection{Current-voltage characteristics}

In the case of the current-voltage characteristics, the perturbative analysis carried out in this work leads to the well known result \cite{ingold:1992}
\beq
I(V) = \frac{\pi e E^2_J}{\hbar}\left[P(2eV) - P(-2EV)\right]\;,
\eeq
with the function
\beq
P(\omega) = \frac{1}{2\pi\hbar} \int^{\infty}_{-\infty} dt \exp\left[{\cal J}(t)+i\omega t\right] \;,
\nonumber
\eeq
and where
\beq
{\cal J}(t) = 2\int^{\infty}_{-\infty} \frac{d\omega}{\omega} \frac{{\rm Re}[Z_t(\omega)]}{R_Q}\frac{e^{-i\omega t}-1}{1 - e^{-\beta\omega}} 
\nonumber
\eeq
is the phase correlation function, with $Z_t(\omega)=Z_L(\omega)+Z_R(\omega)$.
This expression was used to fit the current-voltage characteristic in Ref. \cite{subero_bolometric_2022}.

\subsection{Evidence of insulating behavior in the data of Ref. \cite{subero_bolometric_2022}}

\begin{figure}[t]
\includegraphics[width=\linewidth]{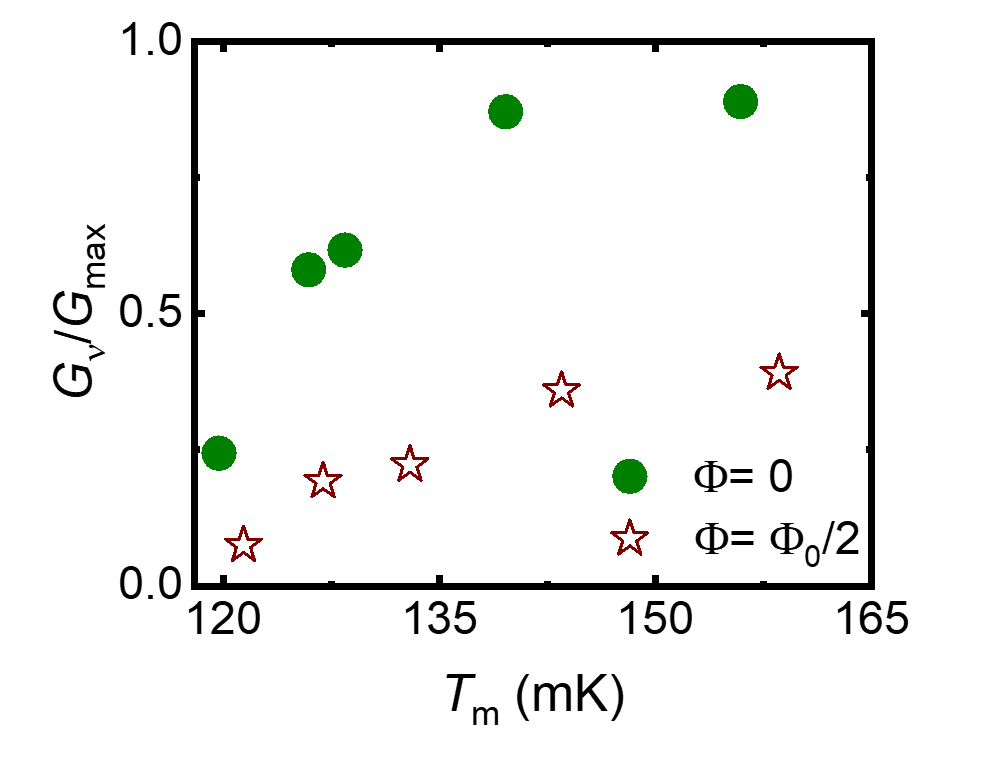}
\caption{Thermal conductance obtained from the data in Ref. \cite{subero_bolometric_2022} normalized to the maximum thermal conductance for a quantum channel. The full dots correspond to zero flux (maximum $E_J$) while the open symbols corresponds to half flux quantum through the SQUID where $E_J \rightarrow 0$.}
\label{conductance-vs-T}
\end{figure}

According to Fig. 3(c) in the main text, a junction in the insulating side of the SB transition should exhibit decreasing heat conductance for $\bar{T} \rightarrow 0$, while it should saturate to a finite value on the superconducting side. The data shown in Fig. \ref{conductance-vs-T}, taken from Ref. \cite{subero_bolometric_2022}, would be thus compatible insulating behavior. A detailed comparison with theory would require, however, measurements in a extended temperature range.

\end{document}